\documentstyle[aps,prd,epsf]{revtex}
\tighten

\begin{document}
\draft

\title{Dynamics of scalar fields in the background of rotating black
holes II: A note on superradiance}
\author{Nils Andersson${}^{(1,2)}$, 
        Pablo Laguna${}^{(3)}$ and
        Philippos Papadopoulos${}^{(4)}$}
\address{
${}^{(1)}$ Institut f\"ur Astronomie und Astrophysik\\ Universit\"at
T\"ubingen, D-72076 T\"ubingen, Germany}

\address{
${}^{(2)}$ Department of Mathematics, University of
Southampton, Southampton, UK }

\address{
${}^{(3)}$ Department of Astronomy \& Astrophysics and \\
Center for Gravitational Physics \& Geometry\\
Penn State University, University Park, PA 16802, USA}

\address{
${}^{(4)}$ Max-Planck-Institut f\"ur Gravitationsphysik\\
Schlaatzweg 1, 14473 Potsdam, Germany}

\date{\today}

\maketitle

\begin{abstract}
We analyze the amplification due to so-called superradiance 
from the scattering
of pulses off rotating black holes as a numerical time
evolution problem.
We consider the ``worst possible case" of scalar field pulses
for which superradiance effects yield amplifications $< 1\%$.
We show that this small effect can be isolated by numerically 
evolving quasi-monochromatic,
modulated pulses with a recently developed Teukolsky code.  
The results show that it is possible 
to study superradiance in the time domain, but only if the initial data
is carefully tuned. This illustrates the intrinsic difficulties 
of detecting superradiance in more general evolution scenarios.
\end{abstract}

\section{Introduction}

In the last few years, we have been involved in the development of
a numerical code
for the time evolution of perturbations of rotating black holes
based on  the Teukolsky
equation \cite{paper1,paper2}. 
There are several motivations for this work. 
One is the desire to
revisit problems that have previously (mainly in the 1970s)
only been approached in the frequency
domain. That is, our goal is to explore the effects of the 
rotation of the black hole
from a ``time-evolution'' point of view. More importantly, our ultimate goal is
to provide a framework that will, once we understand how to
construct astrophysically relevant initial data \cite{ivp}, be used to  
extend the close-limit approximation of head-on black hole collisions
to 
the case of 
inspiral black hole mergers. The working premise in head-on, 
close-limit collisions \cite{price94} is
that the merger can be viewed as perturbations of non-rotating black
holes. In contrast, 
an inspiral close-limit approximation requires perturbations about
a rotating black hole.

Our Teukolsky code project took us first to study the dynamics of
scalar fields in the Kerr geometry \cite{paper1}.
This work mainly concerned the late-time, power-law behaviour of a
scalar perturbation.
The second installment concerned gravitational
perturbations \cite{paper2} 
and discussed the full dynamical response of a black hole
to an external perturbation, namely the quasinormal mode ringing and the subsequent
late-time tails. In Ref.\cite{paper2}, we also 
dealt briefly with superradiance: The anticipated
amplification as certain wavelengths are scattered by
the black hole.  
However, although the results we obtained indicated the presence of
superradiance \cite{paper2}, we feel that our previous analysis 
was not completely satisfactory. 
Hence, the goal of this short paper is to return to the issue of superradiance in
a setting that yields unequivocal evidence for the superradiance phenomenon.

The direct approach to measure superradiance from the time evolution 
of perturbations of rotating black holes is to compute the
energy flux going ``down the hole''. For perturbative fields
that posses well-defined stress-energy tensors (e.g. scalar and electromagnetic
fields), it is possible
to construct such a conserved energy flux \cite{teuk74}.
The case of gravitational perturbations is not that simple \cite{teuk74,hawking72}.
For this reason, we will concentrate our analysis on the ``simple'' case of
scalar perturbations. The price to pay is that superradiant effects in
this case are $< 1\%$ \cite{press72}, thus requiring a highly accurate
evolution code.

For scalar perturbations, the Teukolsky equation in Boyer-Lindquist coordinates reads
\begin{eqnarray}
&&\left[ { (r^2+a^2)^2 \over\Delta} - a^2\sin^2\theta \right] \frac{\partial 
^2 \Phi}{\partial t^2} + \frac{4iMamr}{\Delta} \frac{\partial \Phi}{\partial
t} - \frac{\partial }{\partial r}
\left( \Delta \frac{\partial\Phi}{\partial r} \right) \nonumber \\  
&&-\frac{1}{\sin\theta} \frac{\partial}{\partial \theta}\left(\sin\theta
 \frac{\partial\Phi}{\partial
\theta}\right) -m^2
\left[ \frac{a^2}{\Delta}-\frac{1}{\sin^2\theta}\right]
\Phi =0 \ . 
\label{eq:teuk}\end{eqnarray}
Above, $M$ is the mass of the black hole, $a$ is its angular momentum per unit mass and 
$\Delta \equiv r^2 - 2Mr + a^2$. The two horizons of the black hole
follow from $\Delta = 0$, and
 correspond to $r_\pm = M \pm \sqrt{M^2-a^2}$. Reference to
the azimuthal angle $\varphi$ has been removed 
by assuming that the perturbation has a harmonic dependence $e^{im\varphi}$.

\section{Superradiance in the Frequency Domain}

In the standard approach to solve the Teukolsky equation, one proceeds
via separation of variables. 
For our present purposes, it is sufficient
to note that this essentially corresponds to assuming that i) the
time-dependence of the perturbation is accounted for via Fourier
transformation, and ii) there exists a suitable set of angular
function
that can be used to separate the coordinates $r$ and $\theta$. In the
case of scalar perturbations, the angular functions turn out to be 
standard spheroidal wave-functions \cite{spheroidal}.      
Knowing this, we assume a representation (for each given
integer $m$)
\begin{equation}
\Phi = \int d\omega\, e^{-i\omega t} \sum_{l=0}^\infty R_{lm}(r,\omega) 
S_{lm}(\theta,\omega) \ ,
\label{eq:sep}
\end{equation}
where it should be noted that the angular functions depend explicitly on the
frequency $\omega$; that is, they are intrinsically time-dependent
functions. After a separation of variables of the form given by (\ref{eq:sep}), 
the problem reduces to a
single ordinary differential equation for $R_{lm}(r,\omega)$. This
equation can be written as  
\begin{equation}
{d^2 R_{lm} \over dr_\ast^2} + 
\left[ {K^2 + (2am\omega -a^2\omega^2 - E)\Delta \over (r^2+a^2)^2}
-{dG \over dr_\ast} - G^2 \right] R_{lm} = 0 \ ,
\label{radeq}\end{equation}
where
$K = (r^2 + a^2)\omega - am$, 
$G = r\Delta/(r^2+a^2)^2$, and
the tortoise coordinate $r_\ast$ is defined from 
$ dr_\ast = (r^2 + a^2)/\Delta\, dr$.
The variable $E$ is the angular separation constant. 
In the limiting case
$a\to 0$, it reduces to $l(l+1)$, and for nonzero $a$ it can be obtained
from a power series in $a\omega$ \cite{seidel}. It should be noted that 
$E$ is real valued for real frequencies.     

The physical solution to (\ref{radeq}) is defined by the asymptotic behaviour
\begin{equation}
R_{lm} \sim \left\{ \begin{array}{ll} {\cal T}\, e^{-i(\omega-m\,\omega_+)\,r_\ast} 
\quad &\mbox{as } r\to r_+ \ , \\
{\cal R}\, e^{i\omega r_\ast} +  e^{-i\omega r_\ast} 
\quad &\mbox{as } r\to +\infty \ . \end{array}
\right.
\label{asymp}\end{equation}
where $ \omega_+ \equiv a /2\,M\,r_+$
is the angular velocity of the event horizon. ${\cal T}$ and
${\cal R}$ denote 
the transmission and reflection coefficients, respectively.
Given the prescribed asymptotic behaviour, 
together with that for the complex conjugate of $R_{lm}$ and the fact that
two linearly independent solutions to (\ref{radeq}) must lead
to a constant Wronskian, it is not difficult to show that
\begin{equation} 
( 1 - m\,\omega_+/\omega ) {\cal T}  = 1 - {\cal R}  \ .
\end{equation}
From this result, it is evident that superradiance (${\cal R}>1$)
occurs if
\begin{equation}
\omega < m\,\omega_+ = \frac{m\, a}{ 2\,M\,r_+} \ .
\end{equation}

Alternatively, one can deduce that energy can be extracted from the black 
hole immediately from the boundary condition 
(\ref{asymp}). If $\omega  < m\,\omega_+$, the solution
 $ \propto e^{-i(\omega-m\,\omega_+)\,r_\ast}$,
which behaves as ``ingoing'' into the horizon according to a 
local observer, will in fact correspond to waves coming out of the hole
according to an observer at infinity.
That is, for superradiant frequencies,
one would expect to find energy flowing out from the horizon, cf. \cite{wald}. 
Superradiance is the wave analogue of the standard Penrose process,
and its existence
implies that it would in principle be possible to mine a rotating
black hole for some of its rotational energy.  This may seem
exciting, but it is very unlikely that this effect will play 
a relevant role in any reasonable astrophysical
scenario. Nonetheless, it is an interesting effect that deserves
a close theoretical investigation.     

\begin{figure}[tbh]
\leavevmode
\\
\epsfxsize=0.7\textwidth
\epsfbox{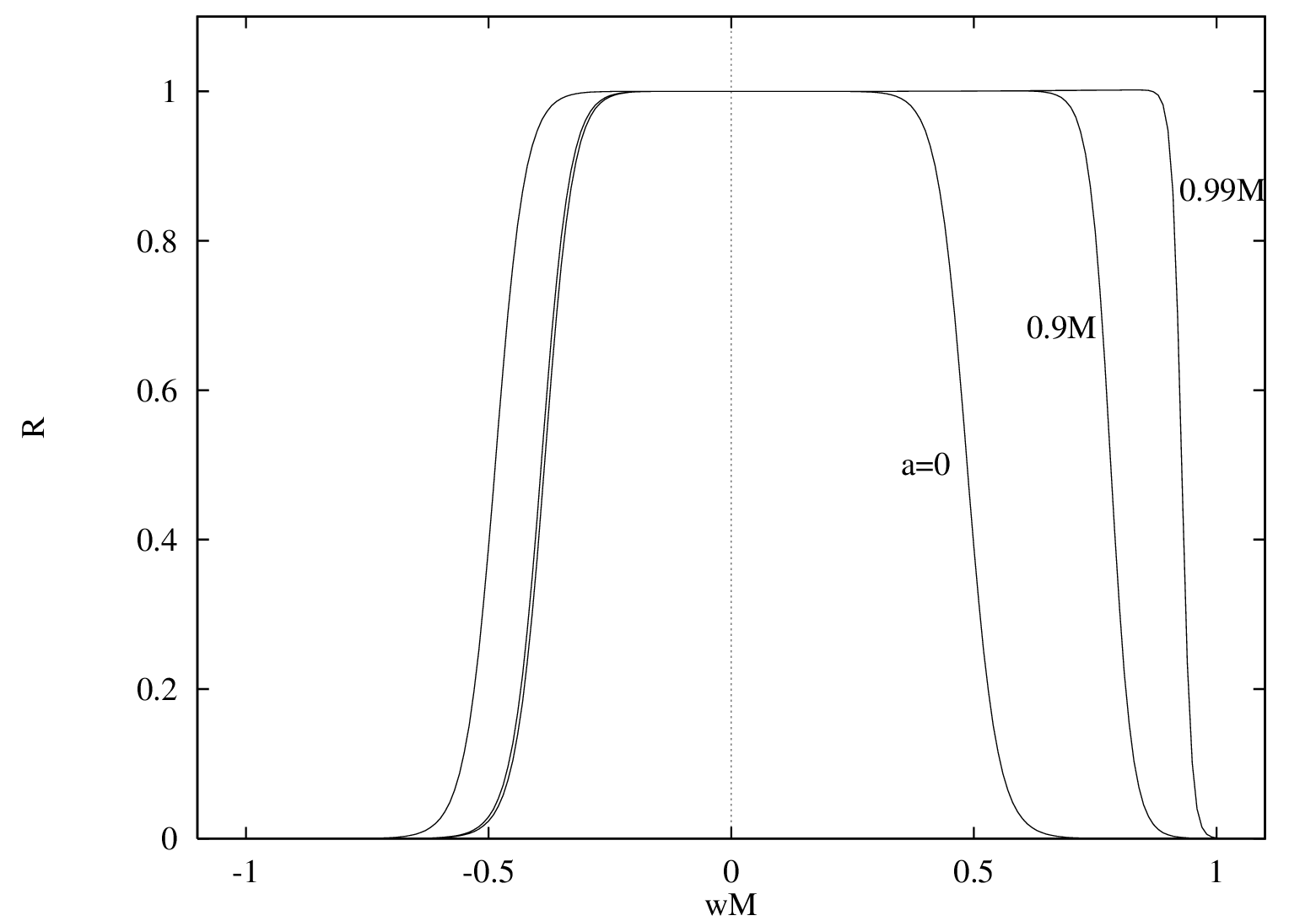}

\epsfxsize=0.7\textwidth
\epsfbox{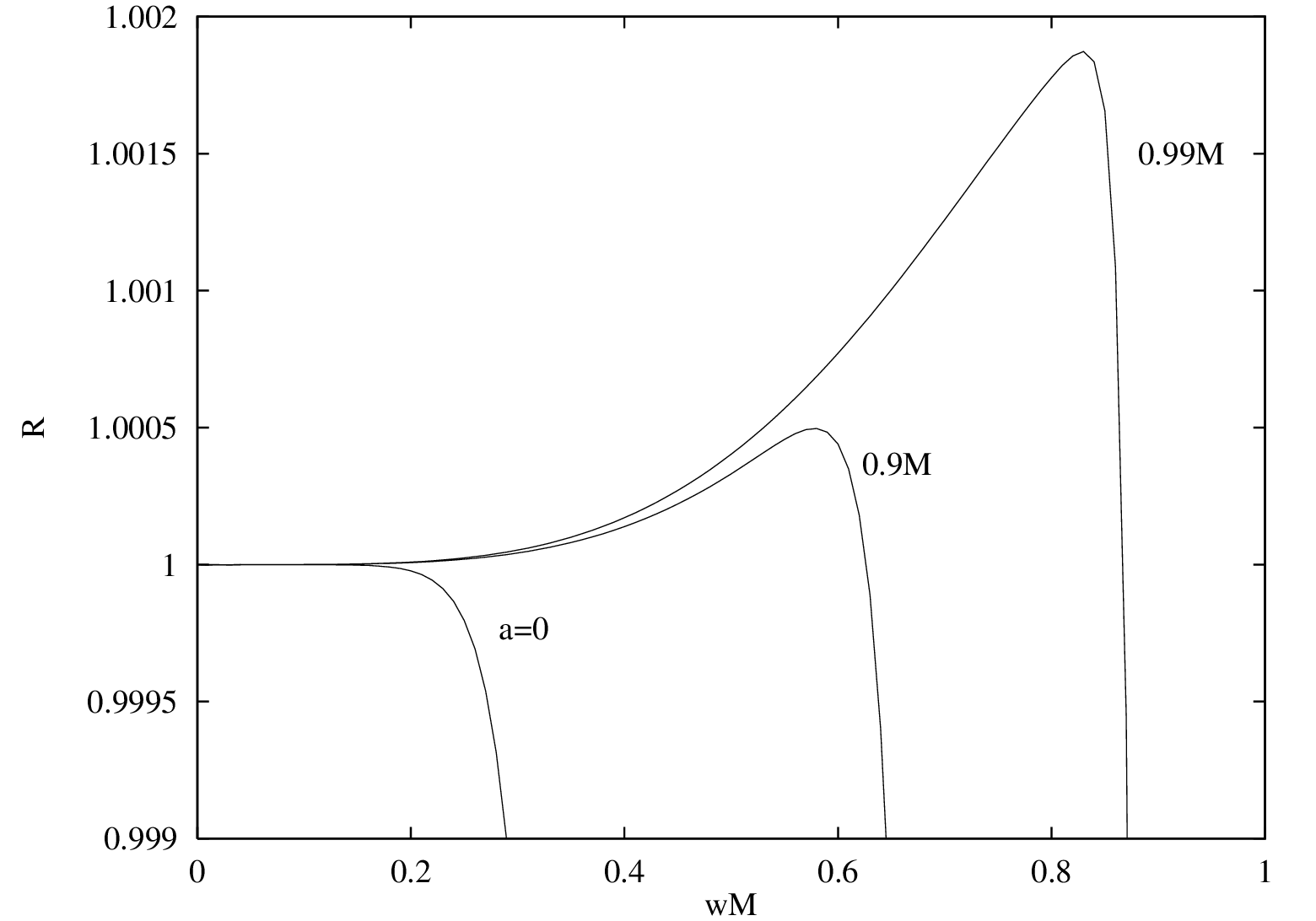}
\caption[fig1]{\label{fig1} Reflection coefficient ($\cal R$) as a function of the
frequency of the wave ($\omega\,M$) for different
values of the angular momentum parameter ($a$) with $l=m=2$. Superradiance is present in the
interval $0 < \omega\,M < m\, a/2\,r_+$, with  $r_+ = M + \sqrt{M^2-a^2}$. 
The bottom panel is a close-up of
this superradiance regime. The maximum observed superradiant amplification 
is $\sim 0.187\%$. 
As $a\to M$, there is
a clearly enhanced reflection of prograde waves ($\omega>0$)
while the overall reflection of retrograde waves ($\omega<0$) 
decreases somewhat. }
\end{figure}

In Fig.~\ref{fig1}, we show a sample of results for the reflection
coefficient in the case when $l=m=2$. These results were obtained by 
a straightforward integration of (\ref{radeq}) and subsequent
extraction of ${\cal R}$. The maximum amplification in this case
is close to 0.2 \%. 
Similar results were obtained by Teukolsky and Press more
than 25 years ago \cite{teuk74,press72}. 
They also considered electromagnetic waves and 
gravitational perturbations and found that
the maximum amplification is 0.3\% for scalar waves, 
4.4\% for electromagnetic waves and as large as
138\% for gravitational waves. 

Given the results in Fig.~\ref{fig1}, it is worth pointing out that
they agree with the standard conclusions regarding the apparent ``size''
of a rotating black hole as seen by different observers. It is
well-known (see for example \cite{bardeen}) that the black hole will
appear larger to a particle moving around it in a retrograde orbit than to a
particle in a prograde orbit. This is illustrated by the fact that the
 unstable circular
photon orbit (at $r=3M$ in the non-rotating case) is located at
$r=4M$ for a retrograde photon, while it lies at $r=M$ for a prograde
photon. The results in Fig.~\ref{fig1} illustrate the same effect:
In our case, we have prograde motion when $\omega/m$ is positive and
retrograde motion when $\omega/m$ is negative. The data in
Fig.~\ref{fig1}
correspond to $m=2$, and the enhanced reflection for positive
frequencies
as $a\to M$ has the 
effect that the black hole ``looks smaller'' to such waves.  
Conversely, the
slightly decreased reflection for negative frequencies leads to the
black hole appearing ``larger'' as $a\to M$.

\section{Superradiance in the Time Domain}

From the discussion in the previous section, it should be 
clear what must be done to
construct initial data sets that would yield 
superradiance with our Teukolsky code. 
Following an idea introduced in Ref.~\cite{paper2}, we
construct superradiant initial data by setting up a
pulse containing mainly frequencies in the interval
$0 < \omega < m\,\omega_+$ (from now on we will assume that $m$ is
positive). Obviously a non-superradiant pulse would be one   
whose main frequency content is outside this window.
Furthermore, it  makes the analysis easier and better suited to
comparisons with frequency domain calculations if
the pulse is ``almost monochromatic.''
To achieve this, we use as initial data an ingoing Gaussian pulse modulated by a
monochromatic wave.
Assuming that this initial pulse is centered far away from the black
hole at $r_\ast=r_o$ and that the
modulation frequency is $\sigma$, we set at $t=0$
\begin{equation}
\Phi \propto e^{-(r_\ast-r_o+t)^2/b^2 - i\sigma (r_\ast-r_o+t)}  \ .
\label{modul}\end{equation}
Thus, the corresponding power spectrum is 
$P(\omega) = P_{max}\, e^{-(\omega-\sigma)^2 b^2 /4}$. 
The modulated pulse leads to a Gaussian
frequency distribution that peaks at a frequency $\omega=\sigma$. 
Given this initial data, and the fact that there is no frequency 
dispersion in the 
Teukolsky equation, we ought to be able to detect
superradiance in our evolutions if
$0 < \sigma < m\,\omega_+$. Conversely, 
if $\sigma > m\,\omega_+$ or $\sigma <0$ we 
should not find any amplification in the scattered waves. 
Finally, it is not enough for the peak of the power spectrum of the 
pulse to lie within the
superradiant frequency window. To maximize the effect, 
we need also to minimize the
``frequency overlap'' of the initial pulse
into the non-superradiant regime. To accomplish this, we have at our disposal
the parameter $b$ that governs the width of the pulse.
If for instance we want $P(m\,\omega_+)/P_{\rm max}=\epsilon$,
we should use
\begin{equation}
b = \frac{2\,\sqrt{\ln (1/\epsilon)}}{m\,\omega_+-\sigma} \ .
\end{equation}

Before showing that the pulse (\ref{modul}) can indeed be used to probe 
superradiance in the time domain, 
a few comments on our Teukolsky code are needed.
The code was described in detail in Ref.~\cite{paper1}, but
there is one specific issue that is  important for the
present study that has not yet been discussed. To avoid numerical
difficulties (cf. the description in \cite{paper1}),
we replace the azimuthal angle
$\varphi$ with the ``ingoing Kerr-coordinate'', which is defined by 
\begin{equation}
\tilde{\varphi} = \varphi + \int {a \over \Delta} dr \ .
\end{equation}
The transformation between the solution $\Phi$ to the standard Teukolsky equation 
and the solution $\tilde{\Phi}$ from our numerical code in terms of $\tilde{\varphi}$
is 
\begin{equation}
\Phi = \tilde{\Phi} \exp{\left[ { im \int {a \over \Delta} dr} \right]} \ .
\label{trans}
\end{equation} 
This is more or less obvious, but it is important to 
notice a few things. First of all, the replacement 
$\varphi \to \tilde{\varphi}$ changes the symmetry of the 
equations. While the original Teukolsky equation (\ref{eq:teuk}) is symmetric under
the change
$(m,\varphi)\to (-m,-\varphi)$, the equation we evolve with our code for $\tilde{\varphi}$ 
is not. That is, while
evolutions for the same Gaussian pulse (unmodulated!) should lead to 
the same emerging scalar waves for $\pm m$ in the original case, this will not
happen when we use $\tilde{\varphi}$.
To ensure that the anticipated symmetries are present in our results,
we have constructed initial data in Boyer-Lindquist
coordinates and then used the transformation (\ref{trans})
to get data in the $\tilde{\varphi}$ coordinate system. 
This is especially important in the case of   
modulated Gaussians since the idea that we could enhance superradiance in the
scattered wave by centering the Gaussian
at a non-zero frequency was based on an analysis in 
Boyer-Lindquist coordinates, and the result of the experiment will
depend on a careful tuning of the initial pulse.

Armed with the above conclusions, we are prepared to discuss our
numerical results. Because we are interested in unveiling the amplification
due to superradiance, we will focus on the energy flux through various
surfaces surrounding the black hole.

Given a spacetime with a time Killing vector $t^a$ (like the  Kerr geometry)
and a perturbation with a 
well-defined stress-energy tensor $T_{ab}$, it is possible to define \cite{teuk74}
a conserved energy flux vector $T^a\,_b\,t^b$. 
The flux of energy across a 3-dimensional time-like hypersurface with
unit normal $r^a$ is then given by
\begin{equation}
dE = T_{ab}\,t^a\, r^b\, dS \ ,
\end{equation}
where $dS$ is the 3-surface element of the hypersurface. 
For a massless scalar field,
\begin{equation}
T_{ab} = \frac{1}{2}(\nabla_a\bar{\Phi}\nabla_b \Phi + \nabla_a \Phi\nabla_b\bar{\Phi})
- \frac{1}{2}g_{ab} \,\nabla_c \Phi \nabla^c \bar{\Phi} \ ,
\end{equation}
with over-bars denoting complex conjugation.
For simplicity, we monitor the flux of energy through $r$ = constant surfaces
in Boyer-Lindquist coordinates. 
This assumption together with $r^a\,r_a = 1$ yield
$r^a = \pm (0,\sqrt{\Delta}/\rho, 0, 0),$
where (as before) $\Delta = r^2 -2Mr+a^2$ and $\rho^2 = r^2+a^2 \cos^2 \theta$.
Furthermore, the time Killing vector in this case reads  $t^a = (1,0,0,0)$, and
the surface element is given explicitly by 
\begin{equation}
dS = \sqrt{-g^{(3)}}\,\sin\theta\, d\theta\, d\varphi\, dt = \sqrt{\Delta}\,\rho
\,\sin\theta\, d\theta\, d\varphi\, dt \ .
\end{equation}
By collecting the above results and noticing that $t^a\,r_a = 0$, it is not
difficult to show that 
\begin{equation}
dE = \pm {1\over 2}(\partial_r\bar{\Phi}\partial_t \Phi +
\partial_r \Phi\partial_t \bar{\Phi}) \Delta \sin\theta\, d\theta
\,d\varphi\, dt \ .
\end{equation}
Finally, integration over $\varphi$ yields
\begin{equation}
dE = \pm \pi\, (\partial_r\bar{\Phi}\partial_t \Phi +
\partial_r \Phi\partial_t \bar{\Phi}) \Delta\, \sin\theta\, d\theta
\,dt \ ,
\end{equation}
which can be rewritten in terms of the tortoise coordinate as
\begin{equation}
dE = \pm \pi (\partial_{r_\ast}\bar{\Phi}\partial_t \Phi +
\partial_{r_\ast} \Phi\partial_t \bar{\Phi}) (r^2+a^2) \sin\theta\, d\theta
\,dt \ .
\end{equation}

In our simulations, the energy flux is monitored 
through two surfaces located at $r_\ast = \pm20M$. 
The outer surface is well away from the black hole while the   
inner one is reasonably  close to the event horizon. The scattering of
waves
by the curved spacetime should be strongest, i.e. most of the
interesting 
physics should have its origin,  in the region
included between these surfaces. 

\begin{figure}[tbh]
\leavevmode
\\
\epsfxsize=0.7\textwidth
\epsfbox{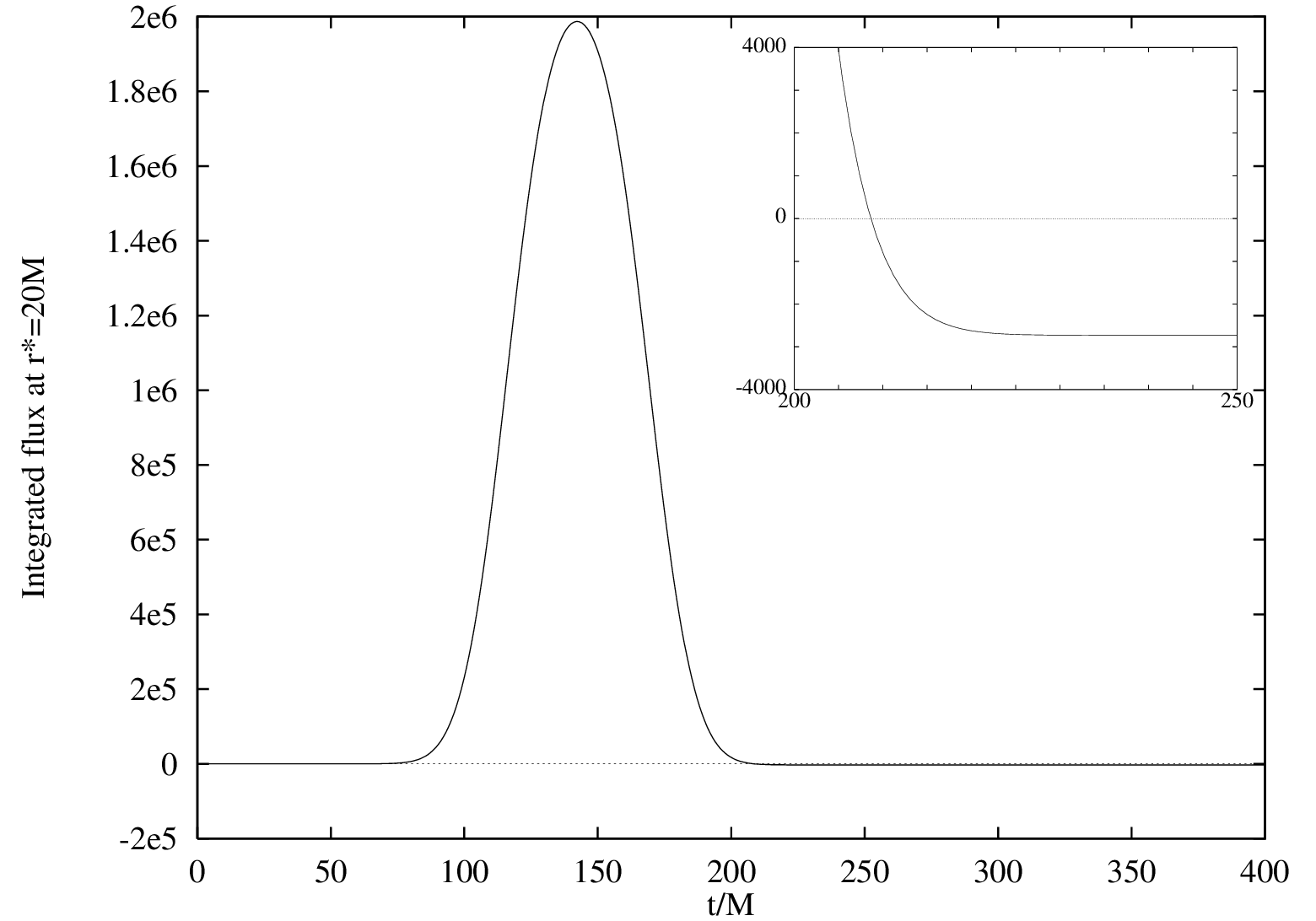}

\epsfxsize=0.7\textwidth
\epsfbox{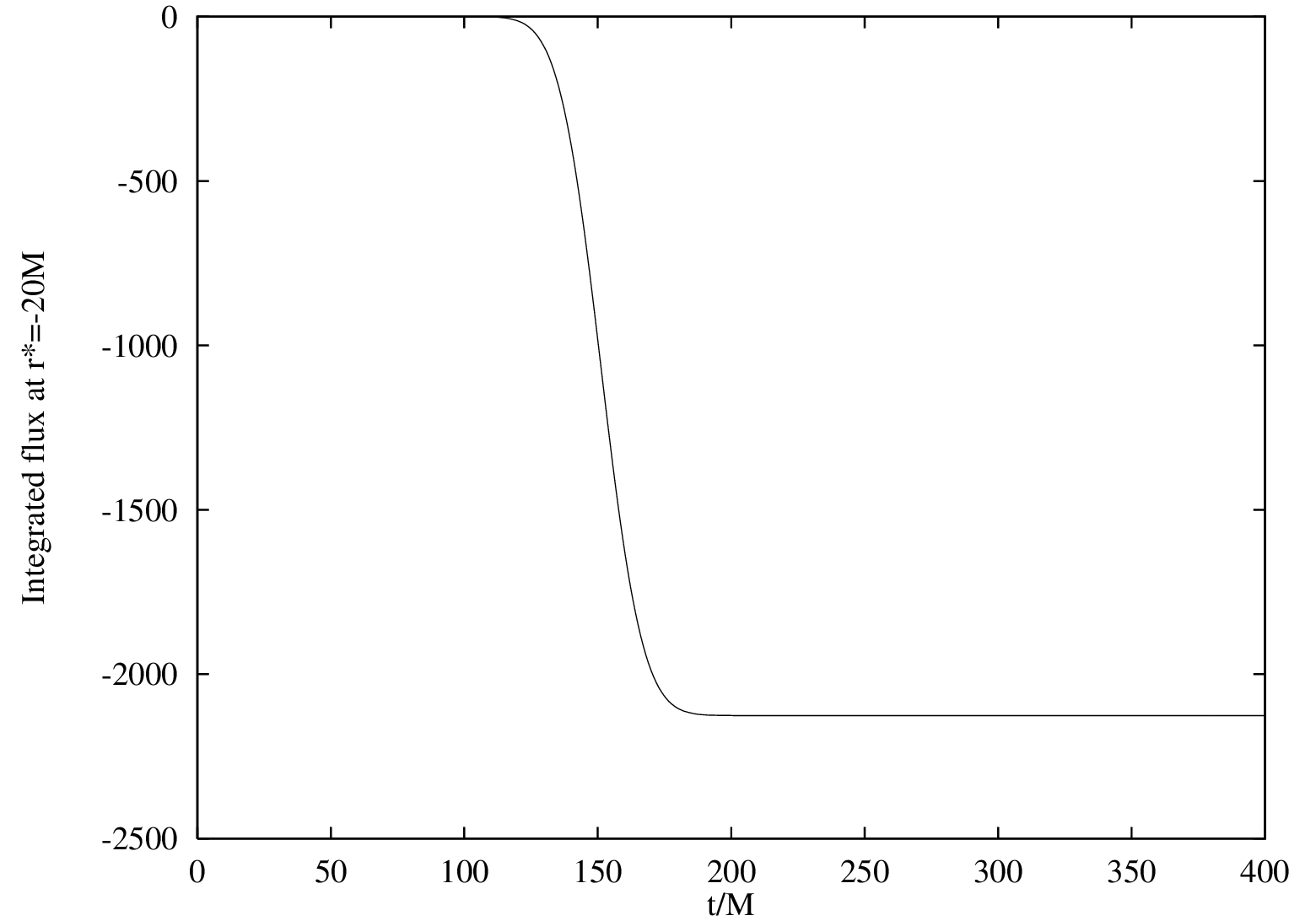}
\caption[fig2]{\label{fig2} An example of a superradiant evolution. We
show results corresponding to a modulated Gaussian pulse (with $\sigma
= 0.8\, m\,\omega_+$, $m=2$, $a=0.99M$ and  $\epsilon = 0.01$, see the main text for
more details). We monitor the integrated energy flux through two surfaces,
one located at  $r_\ast=20M$ and the second at $r_\ast=-20M$. 
 (The normals of these
surfaces are chosen such that energy flowing inwards into-the-hole across the outer
surface is positive.)
Superradiance manifests itself in two ways. At the outer surface (the
upper panel), we see a total
amplification of $\sim 0.14\%$ in the reflected energy. As is clear from the
inset in the upper panel, the reflected energy is larger than what initially
fell onto the black hole. 
At the inner surface (the lower panel), we find that energy mainly
flows out of the horizon. The integrated flux of this energy
corresponds
to $\sim 0.11\%$ of the total energy falling onto the hole, i.e. agrees
well with the amount that amplifies the reflected waves at the outer surface.}
\end{figure}

\begin{figure}[tbh]
\leavevmode
\\
\epsfxsize=0.7\textwidth
\epsfbox{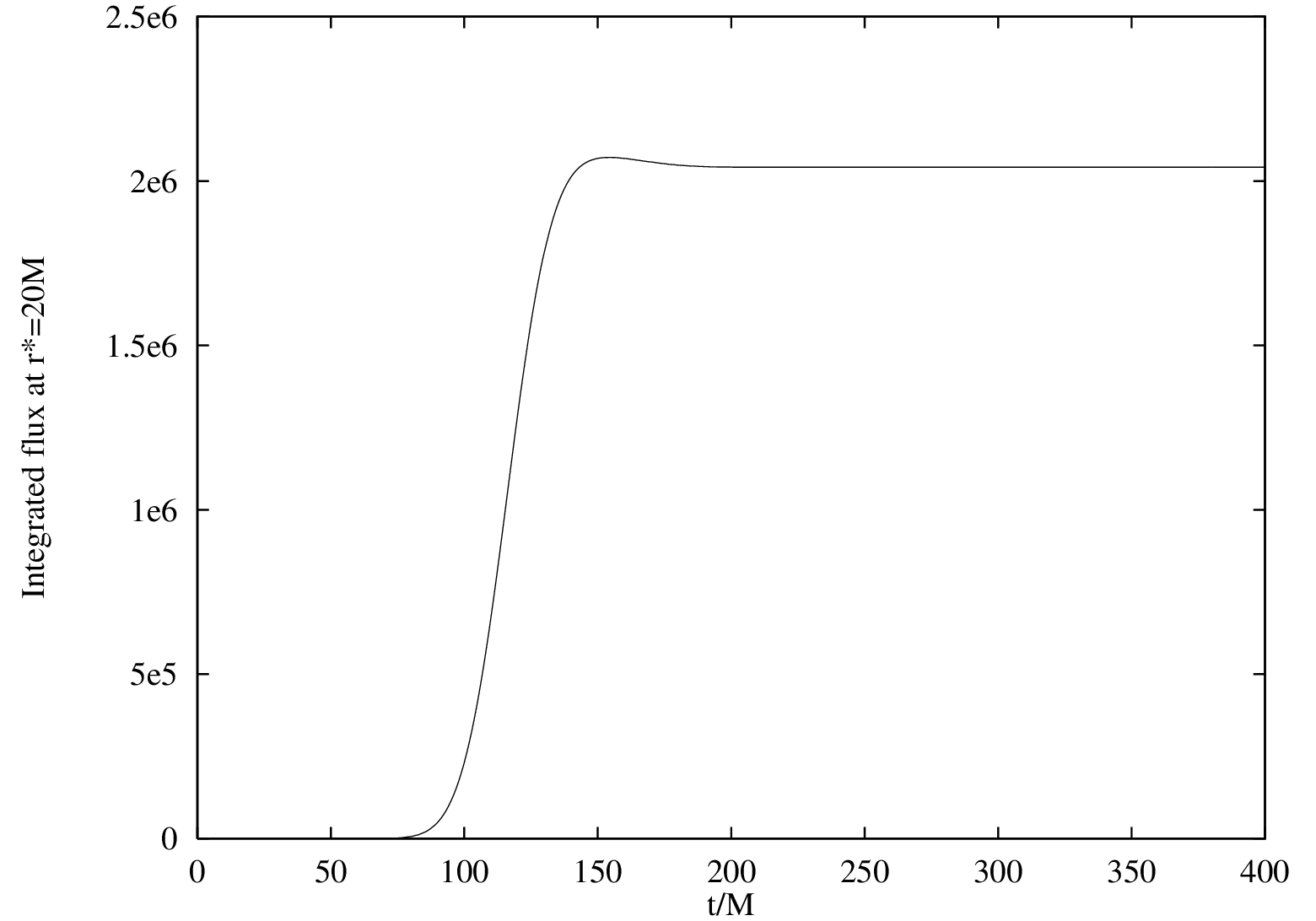}

\epsfxsize=0.7\textwidth
\epsfbox{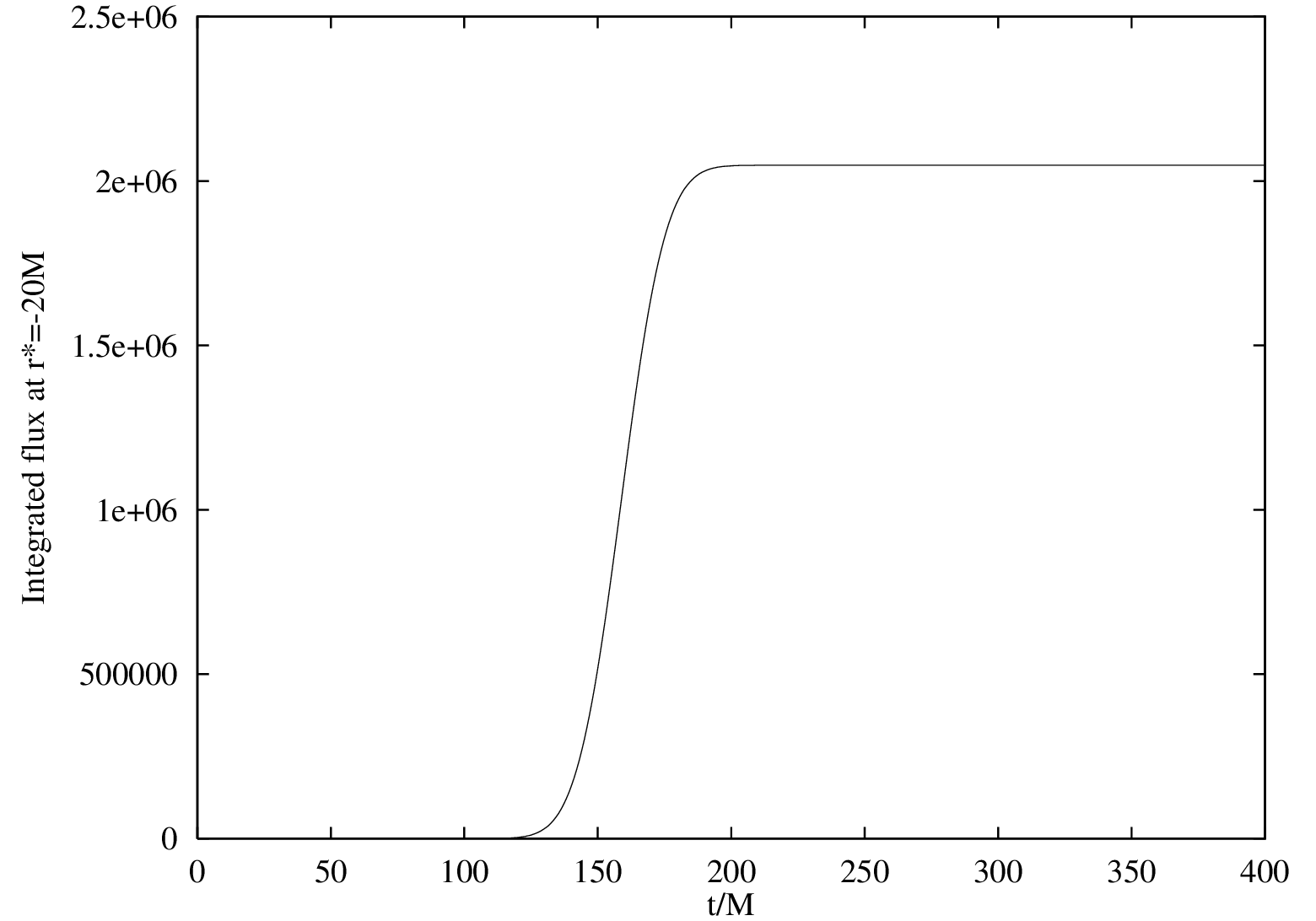}
\caption[fig3]{\label{fig3} An example of a non-superradiant
evolution. The data are similar to that described in Fig.~2, but here the 
modulation frequency of the initial Gaussian is $\sigma
= -0.8\, m\,\omega_+$, and there is no sign of superradiance.   }
\end{figure}

In Figs.~\ref{fig2} and \ref{fig3}, we show the results of
the superradiance ``experiment''. The displayed data are for two
qualitatively different situations. 
Both datasets correspond to  evolutions with $m=2$ and 
a black hole rotation parameter $a=0.99M$. In both cases the pulse
was initially centered around $r_o = 125M$, and 
the angular distribution of the initial data 
was chosen to be the standard spherical harmonic
$Y_2^2(\theta,\varphi)$. 
The first case (see Fig.~\ref{fig2}) shows a situation where one would expect
to see superradiance. We have chosen the modulation frequency
of the impinging Gaussian (see the previous section)
such that $\sigma = 0.8\, m\,\omega_+$, and the width of the
Gaussian
corresponds to $\epsilon = 0.01$. That is, the pulse has its main 
support (in the frequency domain) in the superradiant regime.
The second case (Fig.~\ref{fig3}) corresponds to a Gaussian with the
same
width but now centered around  $\sigma = -0.8\, m\,\omega_+$.

As is obvious from Fig.~\ref{fig1}, the scattering of the two pulses
we have chosen
should be quite different. And not surprisingly, we find that this is
indeed the case. In the superradiant case shown in
Fig.~\ref{fig2}, all 
the initial energy is reflected by the black hole. Superradiance
is distinguished in two ways. By monitoring the energy flowing across
the surface at $r_\ast=20M$, we see that the reflected energy is 
slightly amplified after scattering (cf. the upper panel of
Fig.~\ref{fig2}).
In this specific case, the amplification corresponds to 0.14\%.
It should be compared to the maximum  single frequency
amplification of 0.187\% for $a=0.99M$, deduced from the data in
Fig.~\ref{fig1}  (and also 
the  0.3\% found by Teukolsky and Press \cite{teuk74}).   
That we are seeing superradiance is also clear from the fact  that
energy flows out through the surface at $r_\ast=-20M$  (cf. the lower 
panel of Fig.~\ref{fig2}). The total energy flowing out through the
inner
surface corresponds to a superradiant  amplification of
0.11\%, in reasonable agreement with the result deduced at
the outer surface.

The non-superradiant results, obtained by modulating the Gaussian
with a frequency of  opposite sign to that used in the superradiant
case, are in clear contrast to the superradiant ones. 
In Fig.~\ref{fig3}, there is certainly no amplification of the
reflected wave. In fact, as can be seen from the upper panel of
Fig.~\ref{fig3}, the infalling pulse is almost entirely swallowed by
the black hole. That there would be very little reflection in this
case could, of course, be anticipated by comparing our chosen Gaussian
pulse to the data in Fig.~\ref{fig1}. 

\section{Concluding remarks}

We have designed a numerical experiment that clearly exhibits the presence
of superradiance phenomenon when 
waves of a certain character are scattered by a rotating black hole.
Superradiant effects have previously only been studied in the
frequency domain, and one conclusion that can be drawn from the
present work is that superradiance is perhaps best approached in that way. 
True, we have managed to extract the amplification due to
superradiance in  the ``worst possible case'' of scalar waves,
when the superradiant amplification is expected to be considerably
less than 1\%, but this was mainly due to having at our disposal a conserved
flux. More than anything else this is direct evidence of 
the precision of our evolution code to solve the Teukolsky equation \cite{paper1,paper2}.

The initial data we used to isolate the tiny effect due to
superradiance is required to be ``almost monochromatic'' and thus artificial. 
Our numerical experiment shows that superradiance can play a role in evolutions
when the scattered pulse has support only in a restricted frequency domain. 
This is undoubtedly an interesting illustration, but
what about superradiance in more general cases? It seems to us that the effect
is easiest to  isolate if one monitors different frequencies separately, 
i.e. works in the frequency domain. 
The main reason for this is that an amplification   
of a reflected signal with increasing $a$ is not
in itself an indication of superradiance. The results shown in
Fig.~\ref{fig1} indicate that one would generally expect enhanced
reflection of prograde moving waves as $a\to M$. This effect is
likely to overwhelm the actual ``amplification'' of certain superradiant
frequencies in an evolution of general initial data. This is certainly
true for scalar waves, and since the maximum amplification of
impinging electromagnetic waves is only a few percent \cite{teuk74}
the 
conclusion
should hold also in that case. However,  
the possibility that superradiance may play a distinctive role in an 
``astrophysical'' evolution for gravitational waves cannot be ruled out. 
For gravitational waves, 
the amplitude of certain frequencies should be amplified
by more than a factor of two \cite{teuk74}. A detailed study of that
case could provide interesting results, but the present results
for scalar field prompts us to proceed with caution. We have learned
that the initial data requires careful tuning in order that
superradiance be observed. For general initial data, absorption of the 
non-superradiant frequencies will typically make the amplification due to
superradiance
difficult to distinguish. Moreover, 
superradiance should not be confused with the competing effect that
the ``size'' of the black hole changes with the rate of rotation.
As we have seen, this effect will generally lead to a much enhanced
reflection of prograde waves
 which may confuse an attempt to distinguish superradiance.

\section{Acknowledgments}

We thank William Krivan for helpful discussions. This work was partially supported by
NSF grants PHY 96-01413, 93-57219 (NYI) to PL.

\end{document}